# An Analytical Approach to Document Clustering Based on Internal Criterion Function

Alok Ranjan
Department of Information Technology
ABV-IIITM
Gwalior, India

Eatesh Kandpal
Department of Information Technology
ABV-IIITM
Gwalior, India

Harish Verma
Department of Information Technology
ABV-IIITM
Gwalior, India

Joydip Dhar
Department of Applied Sciences
ABV-IIITM
Gwalior, India

*Abstract*—Fast and high quality document clustering is an important task in organizing information, search engine results obtaining from user query, enhancing web crawling and information retrieval. With the large amount of data available and with a goal of creating good quality clusters, a variety of algorithms have been developed having quality-complexity trade-offs. Among these, some algorithms seek to minimize the computational complexity using certain criterion functions which are defined for the whole set of clustering solution. In this paper, we are proposing a novel document clustering algorithm based on an internal criterion function. Most commonly used partitioning clustering algorithms (e.g. k-means) have some drawbacks as they suffer from local optimum solutions and creation of empty clusters as a clustering solution. The proposed algorithm usually does not suffer from these problems and converge to a global optimum, its performance enhances with the increase in number of clusters. We have checked our algorithm against three different datasets for four different values of k (required number of clusters).

*Keywords*—*Document clustering; partitioning clustering algorithm; criterion function; global optimization*

## I. INTRODUCTION

Developing an efficient and accurate clustering algorithm has been one of the most favorite areas of research in various scientific fields. Various algorithms have been developed over a period of years [2, 3, 4, 5]. These algorithms can be broadly classified into agglomerative [6, 7, 8] or partitioning [9] approaches based on the methodology used or into hierarchical or non-hierarchical solutions based on the structure of solution obtained.

Hierarchical solutions are those which are in the form of a tree called dendograms [15], which can be obtained by using agglomerative algorithms, in which, first each object is assigned to its own cluster and then pair of clusters are repeatedly joined until a certain stopping condition is not satisfied. On the other hand , partitioning algorithms such as k-means [5], k-medoids [5], graph-partioning-based [5] consider whole data as a single cluster and then find clustering solution by bisecting or partitioning it into number of predetermined classes. However, a repeated application of partitioning application can give a hierarchical clustering solution.

There always involves tradeoffs between a clustering solution quality and complexity of algorithm. Various researchers have shown that partitioning algorithms in terms of clustering quality are inferior in comparison to agglomerative algorithms [10]. However, for large document datasets they perform better because of small complexity involved [10, 11].

Partitioning algorithms work using a particular criterion function with the prime aim to optimize it, which determines the quality of clustering solution involved. In [12, 13] seven criterion functions are described categorized into internal, external and hybrid criterion functions. The Best way to optimize these criterion functions in partitioning algorithmic approach is to use greedy approach as in k-means. However the solution obtained may be sub-optimal because many a times these algorithms converge to a local-minima or maxima. Probability of getting good quality clusters depends on the initial clustering solution [1]. We have used an internal criterion function and proposed a novel algorithm for initial clustering based on partitioning clustering algorithm. In particular we have compared our approach with the approach described in [1] and implementation results show that our approach performs better then the above method.

## II. Basics

In this paper documents have been represented using a vector-space model [14]. This model visualizes each document, d as a vector in the term-space or more in more precise way each document d is represented by a term-frequency (T-F) vector.

$$d_{tf} = (tf_1, tf_2,................, tf_m),$$

where $tf_i$ denotes the frequency of the $i_{th}$ term in the document. In particular we have used a term-inverse document frequency (tf-idf) term weighing model [14]. This model works better when some terms appearing more frequently in documents having little discrimination power need to be de-emphasized. Value of idf is given by log (N $/df_i$), where N is the total number of documents and $df_i$ is the number of documents that contain the $i_{th}$ term.

$$d_{tf-idf} = (tf_1 \log(N/df_1), tf_2 \log(N/df_2),\dots\dots,$$
$$tf_m \log(N/df_m)).$$

As the documents are of varying length, the document vectors are normalized thus rendering them of unit length ($|d_{tf-idf}|$=1).

In order to compare the document vectors, certain similarity measures have been proposed. One of them is cosine function [14] as follows

$$\text{Cos}(d_i, d_j) = \frac{d_i^t d_j}{||d_i|| \, ||d_j||}$$

Where, $d_i$, $d_j$ are the two documents under consideration, $\| d_i \|$ and $\|d_j\|$ are the lengths of vector $d_i$ and $d_j$ respectively. This formula, owing to the fact that $d_i$ and $d_j$ are normalized vectors, converges into

$$\text{Cos}(d_i, d_j) = d_i^t d_j$$

.The other measure is based on Euclidean distance, given by

$$\text{Dis}(d_i, d_j) = \sqrt{(d_i - d_j)^t (d_i - d_j)} = \| d_i - d_j \|.$$

Let A be the set of document vectors, the centroid vector $C_A$ is defined to be

$$C_A = D_A / |A|$$

where, $D_A$ represents composite vector given by $\sum_{d \in A} d$

## III. DOCUMENT CLUSTERING

Clustering is an unsupervised machine learning technique. Given a set $A_n$ of documents, we define clustering as a technique to group similar documents together without the prior knowledge of group definition. Thus, we are interested in finding k smaller subsets $S_i$ (i = 1, 2,.........k) of $A_n$ such that documents in same set are more similar to each other while documents in different sets are more dissimilar. Moreover, our aim is to find the clustering solution in the context of internal criterion function.

### A. Internal Criterion Function

Internal criterion functions account for finding clustering solution by optimizing a criterion function defined over documents which are in same set only and doesn't consider the effect of documents in different sets.

The criterion function we have chosen for our study attempts to maximize the similarity of a document within a cluster with its cluster centroid [11]. Mathematically it is expressed as

Maximize T = $\sum_{r=1} \sum_{d \in S} Cos(d_i, C_r)$

Where, $d_i$ is the $i_{th}$ document and $C_r$ is the centroid of the $r_{th}$ cluster.

## IV. ALGORITHM DESCRIPTION

Our algorithm is basically a greedy one, unlike other partitioning algorithm (e.g. k-means) it generally does not converge to a local minimum.

Our algorithm consists of mainly two phases (i) initial clustering (ii) refinement.

### A. Initial clustering

This phase consists of determining initial clustering solution which is further refined in refinement phase, with the assumption

In this phase of algorithm, our aim is to select K documents, hereafter called seeds, which will be used as initial centroid of K clusters required.

We select the document which has minimum sum of squared distances from the previously selected documents. In the process we get the document having largest minimum distance from previously selected documents, i.e., document which is not in the neighborhood of currently present documents.

Let at some time we have m documents in the selected list, we check the sum S = $\sum_{i=1}^{k} \left(Dist(d_i, a)\right)^2$ for all documents a in set A, where set A contains the documents having largest sum of distances from previously selected m documents, and finally the document having minimum value of S, is selected as the (m+1)th document. We continue this operation until we have K documents in the selected list.

1) *Algorithm:*

*Step1:* DIST ← adjacency matrix of document vectors

*Step2:* R ← regulating parameter

*Step3:* LIST ← set of document vectors

*Step4:* N ← number of document vectors

*Step5:* K ← number of clusters required

*Step6:* ARR_SEEDS ← list of seeds initially empty

*Step7:* Add a randomly selected document to ARR_SEEDS

*Step8:* Add to ARR_SEEDS a new document farthest from the residing document in ARR_SEEDS

*Step9:* Repeat steps 10 to 13 while ARR_SEEDS has less than K elements

*Step10:* STORE ← set of pair ( sum of distances of all current seeds from each document, document ID)

*Step11:* Add in STORE the pair(sum of distances of all current seeds from each document, document ID)

*Step12:* Repeat Step 13 R times

*Step13:* Add to ARR_SEEDS the document having least sum of squared distances from available seeds

*Step14:* Repeat 15 and 16 for all remaining documents

*Step15:* Select a document

*Step16:* Assign selected document to the cluster corresponding to its nearest seed

2) *Description:* The Algorithm begins with putting up of a randomly selected document into an empty list of seeds named ARR_SEEDS. We define a seed as a document which represents a cluster. Thus we aim to choose K seeds each representing a single cluster. The most distant document from the formerly selected seed is again inserted into ARR_SEEDS. After the selection of two initial seeds, others are to be selected through an iterative process where in each iteration we put all the documents in descending order of their sum of distance from the currently residing seeds in ARR_SEEDS and then from the ordered list we take top R (regulating variable which is to be decided by the total number of documents, the distribution of the clusters in K-dimensional space and the total number of clusters K) documents to find the document having minimum sum of squared distances from the currently residing seeds in the list, the document thus found is added immediately into ARR_SEEDS and more iterations follow until number of seeds reach K. The variable R is a regulating variable which is to be decided by the total number of documents, the distribution of the clusters in K-dimensional space and the total number of clusters K.

Now we have K seeds in ARR_SEEDS each representing a cluster. For the remaining N-K documents, each document is assigned to the cluster corresponding to its nearest seed.

### B. Refinement

The refinement phase consists of many iterations. In each iteration all the documents are visited in random order, a document $d_i$ is selected from a cluster and it is moved to other k-1 clusters so as to optimize the value of criterion function. If a move leads to an improvement in the criterion function value then $d_i$ is moved to that cluster. A soon as all the documents are visited an iteration ends. If in an iteration there are no documents remaining, such that their movement leads to improvement in the criterion function, the refinement phase ends.

1) *Algorithm:*

*Step1:* S ← Set of clusters obtained from initial clustering

*Step2:* Repeat steps 3 to 9 until even a single document moved between clusters

*Step3:* Unmark all documents

*Step4:* Repeat steps 5 to 9 while each document is not marked

*Step5:* Select a random document X from S

*Step6:* If X is not marked , perform Steps 7 to 9

*Step7:* Mark X

*Step8:* Search cluster C in T in which X lies

*Step9:* Move X to any cluster other than C by which the overall criterion function value of S goes down. If no such cluster exists don't move X.

## V. IMPLEMENTATION DETAILS

To test our algorithm we have coded it and the older one in Java Programming language. The rest of this section describes about the input dataset and cluster quality metric entropy which we have used in our paper.

### A. Input Dataset

For testing purpose we have used both a synthetic dataset and a real dataset.

1) *Synthetic Dataset*

This dataset contains a total 15 classes from different books and articles related to different fields such as art, philosophy, religion, politics etc. The description is as follows.

TABLE 1        SYNTHETIC DATASET

| Class label | Number of documents | Class label | Number of documents |
|---|---|---|---|
| Architecture | 100 | History | 100 |
| Art | 100 | Mathematics | 100 |
| Business | 100 | Medical | 100 |
| Crime | 100 | Politics | 100 |
| Economics | 100 | Sports | 100 |
| Engineering | 100 | Spiritualism | 100 |
| Geography | 100 | Terrorism | 100 |
| Greek Mythology | 100 | | |

### 2) Real Dataset

It consists of two datasets namely re0 and re1 [16]

TABLE 2        REAL DATASET

| Data | Source | Number of documents | Number of classes |
|---|---|---|---|
| re0 | Reuters-21578 | 1504 | 13 |
| re1 | Reuters-21578 | 1657 | 25 |

### B. Entropy

Entropy measure uses the class label of a document assigned to a cluster for determining the cluster quality. Entropy gives us the information about the distribution of documents from various classes within each cluster. An ideal clustering solution is the one in which all the documents of a cluster belong to a single class. In this case the entropy will be zero. Thus, the smaller value of entropy denotes a better clustering solution.

Given a particular cluster Sr of size Nr, the entropy [1] of this cluster is defined to be

$$E(S_r) = -\frac{1}{\log q} \sum_{i=1}^{q} \frac{N_r^i}{N_r} \log(\frac{N_r^i}{N_r})$$

where q is the number of classes available in the dataset, and $N_r^i$ is the number of documents belonging to the $i_{th}$ class that were assigned to the $r_{th}$ cluster. The total entropy will be given by the following equation

$$Entropy = \sum_{r=1}^{k} \frac{N_r}{N} E(S_r)$$

### VI.        RESULTS

In this paper we used entropy measure for determine the quality of clustering solution obtained. Entropy value for a particular k-way clustering is calculated by taking the average of entropies obtained from ten executions. Then these values are plotted against four different values of k, i.e., number of clusters. Experimental results are shown in the form of graphs [see Figure 1-3]. The first graph is obtained using the synthetic dataset having 15 classes. The second one is obtained using dataset re0 [16] and the third one is obtained using dataset re1 [16]. The results reveals that the entropy values obtained using our novel approach is always smaller, hence it is better then [1]. Also it is obvious from the graphs that the value of entropy decreases with the increase in the number of clusters as expected.

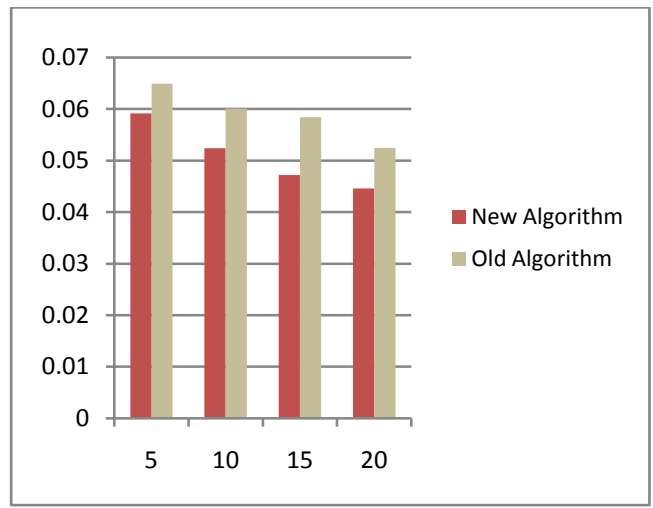

Figure 1.   Variation of entropy Vs number of clusters for synthetic dataset (# of classes 15)

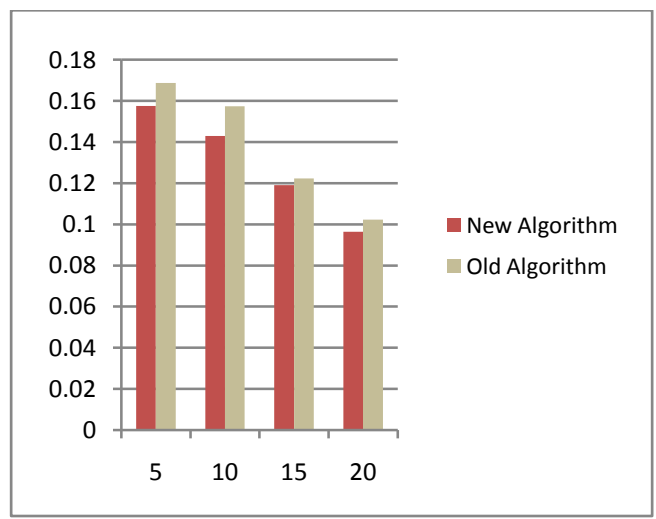

Figure 2.   Variation of entropy Vs number of clusters for dataset re0 (# of classes 13)

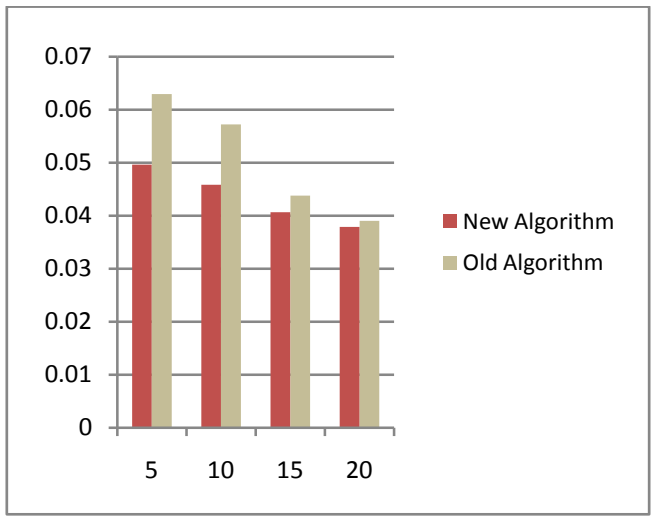

Figure 3.    Variation of entropy Vs number of clusters for dataset re1 (# of classes 25)

## VII.    CONCLUSIONS

In this paper we have successfully proposed and tested a new algorithm that can be used for accurate document clustering. We know that the most of the previous algorithms have a relatively greater probability to trap in local optimal solution. Unlike them this algorithm has a very little chance to trap in local optimal solution, and hence it converges to a global optimal solution. In this algorithm, we have used a completely new analytical approach for initial clustering which refines result and it gets even more refined after the completion of refinement process. The performance of the algorithm enhances with the increase in the number of clusters.